\documentclass[journal=jctc,manuscript=article]{achemso}

\usepackage[version=3]{mhchem} 
\usepackage{amsmath}
\usepackage{multirow}
\usepackage{makecell}
\usepackage[sort&compress,numbers]{natbib}
\usepackage{amsfonts}
\usepackage{multicol}
\usepackage{mciteplus}
\usepackage{epsfig}
\usepackage{float}
\usepackage{fancyhdr}
\usepackage{color}
\usepackage{fixltx2e}
\usepackage{epsfig}
\usepackage{epstopdf}
\usepackage{subcaption}
\usepackage{algorithm}
\usepackage[noend]{algpseudocode}

\newcommand{\erfc}{\mathrm{erfc}\,}

\author{Dor Gabay}
\affiliation{Department of Physical Electronics, Tel-Aviv University, Tel-Aviv 69978, Israel}
\author{Xueyang Wang}
\affiliation{Department of Electrical and Computer Engineering, University of California, San Diego, La Jolla, CA 92093, USA}
\author{Vitaly Lomakin}
\affiliation{Department of Electrical and Computer Engineering, University of California, San Diego, La Jolla, CA 92093, USA}
\author{Amir Boag}
\affiliation{Department of Physical Electronics, Tel-Aviv University, Tel-Aviv 69978, Israel}
\author{Manish Jain}
\affiliation{Department of Physics, Indian Institute of Science, Bangalore 560 012, India}
\author{Amir Natan}
\affiliation{Department of Physical Electronics, Tel-Aviv University, Tel-Aviv 69978, Israel}
\altaffiliation{The Sackler Center for Computational Molecular and Materials Science, Tel-Aviv University, Tel-Aviv 69978, Israel }
\email{amirnatan@post.tau.ac.il}

\title[An \textsf{achemso} demo]
  {Size dependent electronic properties of silicon quantum dots - an analysis with hybrid, screened hybrid and local density functional theory}

\abbreviations{PL,DFT,HOMO,LUMO,GPU,FFT,QD.DOS}

\begin{document}

\begin{abstract}
We use an efficient projection scheme for the Fock operator to analyze the size dependence of silicon quantum dots (QDs) electronic properties. We compare the behavior of hybrid, screened hybrid and local density functionals as a function of the dot size up to $\sim$800 silicon atoms and volume of up to $\sim$20nm$^3$. This allows comparing the calculations of hybrid and screened hybrid functionals to experimental results over a wide range of QD sizes. We demonstrate the size dependent behavior of the band gap, density of states, ionization potential and HOMO level shift after ionization. Those results are compared to experiment and to other theoretical approaches, such as tight-binding, empirical pseudopotentials, TDDFT and GW. 
\end{abstract}

\section{Introduction}
Quantum dots (QDs) form a class of nanometer scale materials that present an efficient way to tune the electronic and optical properties of materials by controlling their size\cite{alivisatos1996semiconductor,yoffe2001semiconductor,chan2002luminescent}. In particular, the optical band gap and electronic density of states are affected by quantum confinement\cite{alivisatos1996semiconductor,yoffe2001semiconductor,chan2002luminescent}. Silicon based QDs are especially interesting as silicon is a highly abundant element with a wide use in electronics, photovoltaics, and many other fields. They were therefore the subject of both experimental\cite{furukawa1988quantum,wolkin1999electronic,park2001band,park2001quantum,belyakov2008silicon,sychugov2016single} and theoretical\cite{chelikowsky2009algorithms,wilson2014shape,delley1995size,zhou2003electronic,belyakov2008silicon,sychugov2016single,reboredo1999excitonic,ougut1997quantum} research. 

Even a small QD can have hundreds of atoms and hence present a challenge for full quantum calculations. Semi-empirical quantum methods such as tight-binding\cite{trani2011silicon} (TB) and empirical pseudopotentials\cite{reboredo1999excitonic} (PP) were successfully applied to silicon QDs. The advantage of such methods is that they can easily treat much larger dots (up to million atoms), in addition, their parameters can be tuned to achieve high agreement with experiment. 

Desnity Functional Theory (DFT)\cite{koch2000chemist} is a first principles quantum approach that offers a reasonable balance between computational cost and level of approximation. DFT with purely local functionals such as the Local Density Approximation (LDA)\cite{koch2000chemist} was used to calculate large silicon quantum dots up to 10,000 atoms\cite{chelikowsky2009algorithms}, however, the LDA functional approximation is known to underestimate the band gap. A generally successful approach for the optical gap calculation is to use time-dependent DFT (TDDFT)\cite{runge1984density} but this method is more computationally expensive and calculations for silicon QDs were performed up to $\sim$150 atoms.\cite{vasiliev2001ab,garoufalis2001high}

The use of hybrid\cite{koch2000chemist}, screened
hybrid\cite{heyd2003hybrid,heyd2004assessment,heyd2004efficient} and range
separated
functionals\cite{toulouse2004long,baer2005density,livshits2007well,kronik2012excitation,kronik2013molsolids,kronik2015tddft}
is becoming one the methods of choice in DFT calculations to achieve reliable results for the
electronic structure of molecules and solids. Screened hybrids such as the
HSE\cite{heyd2003hybrid,heyd2004assessment,heyd2004efficient} functional were shown to give a reliable prediction of
electronic structure for both metallic and insulating materials, thus,
outperforming both purely local methods such as LDA and hybrid functional methods. Recently, the method of optimally-tuned
range separated functionals has shown great success in predicting the correct
gap for a large variety of molecules and molecular
crystals\cite{kronik2012excitation,kronik2013molsolids,kronik2015tddft}. 

A computational barrier for such hybrid and screened hybrid calculations is the need to calculate
the non-local Fock exchange operator for systems with many electrons. We have
recently shown how the use of projection operators in the real-space method can
dramatically reduce the computational time for Hartree-Fock and hybrid
functionals\cite{boffi2016efficient}. In the current work we demonstrate the
implementation of screened hybrid functionals together with a faster, FFT based
Poisson solver\cite{gabay2017kernel}, with an optional GPU implementation that
can give a significant {\it additional} acceleration. These implementations,
enable us to analyze hydrogen passivated silicon QDs with 30-800 silicon atoms (1200 atoms overall), and to examine the
performance of purely local, hybrid, and screened hybrid functionals for those
systems.

This manuscript is organized as follows - we first
review the real-space implementation for the Fock projection scheme and the
screening implementation for the HSE functional. We then demonstrate the effect
of quantum dot size on the Density of States (DOS) and band gap with the HSE06\cite{krukau2006influence},
PBE0\cite{adamo1999toward} and the LDA functionals. We compare our band gap results to photoluminescence (PL) experiments and other theoretical calculations. An additional behavior that we
analyze with HSE06 and LDA is the ionization potential (IP) and the HOMO (Highest
Occupied Molecular Orbital) level shift after ionization. We show that this shift does not depend much on the level of theory that is used and explain this by electrostatic arguments.

\section{Projection
scheme and screening implementation}
We have used the projection scheme for the Fock operator as described in Boffi et al.\cite{boffi2016efficient} within the PARSEC pseudopotential code\cite{Chelikowsky1994_PRL,Chelikowsky1994,Kronik2006} and repeat it very shortly for completeness. The Fock operator can be described by:

\begin{equation}
\hat{K}\psi_\sigma(r)=\sum_{n=1}^{N} \left(\int d\mathbf{r}_1 \frac{\varphi^*_{n,\sigma}(\mathbf{r}_1)\psi_\sigma(\mathbf{r}_1)}{|\mathbf{r}-\mathbf{r}_1|}\right) \varphi_{n,\sigma}(\mathbf{r})\equiv \sum_{n=1}^{N} V_{n,\psi,\sigma}(\mathbf{r})\varphi_{n\sigma}(\mathbf{r}).
\label{eq:fock_operator}
\end{equation}

The use of the operator in Eq. \ref{eq:fock_operator} for an arbitrary orbital is too expensive and so we have used the projection approximation\cite{boffi2016efficient,Duchemin2010}:
\begin{equation}
\mathbf{\tilde{\hat{K}}} \equiv \mathbf{\hat{K} \cdot \hat{P}_M + \hat{P}_M\cdot\hat{K}-\hat{P}_M\cdot\hat{K}\cdot\hat{P}_M}
\end{equation}
where $\mathbf{\hat{P}_M}\equiv \sum_{n=1}^{N_\sigma+M} |\varphi_{n,\sigma}\rangle\langle \varphi_{n,\sigma}| $ , is the projection operaror, projecting over occupied and first $M$ un-occupied states. The details of this scheme are described in Boffi et al.\cite{boffi2016efficient}, similar approaches have been reported by Duchemin and Gygi\cite{Duchemin2010} and others\cite{lin2016adaptive,hu2017adaptively} in the plane waves basis. The calculation of all Poisson integrals in Eq. \ref{eq:fock_operator} is calculated with an FFT based solver for isolated boundary conditions with a numerically optimized kernel as described in Gabay et al. \cite{gabay2017kernel,zuzovski2015}. The combination of projection with the numerically optimized kernel allows us to study large systems which were prohibitively expensive in the traditional implementation.

\subsection{Screening implementation - HSE}
The Heyd-Scuseria-Ernzerhof (HSE) functional\cite{heyd2003hybrid} includes a fraction of screened Fock exchange and requires to add screening to the operator in Eq~\ref{eq:fock_operator}. Formally one can write the energy for the HSE functional as:\cite{heyd2003hybrid}

\begin{equation}
E_{XC}^{HSE}(\omega)=a\cdot E_{X}^{HF,SR}(\omega)-a\cdot E_X^{PBE,SR}(\omega)+E_{XC}^{PBE}
\label{eq:HSE_energy}
\end{equation}

where SR stands for short range, $a$ is 0.25 and $\omega=0.2$\AA$^{-1}$ for HSE06. The short range part of the PBE exchange is implemented as in Heyd et al.\cite{heyd2003hybrid} and we write the screened version of Eq. \ref{eq:fock_operator} as:

\begin{align}
\hat{K}^{SR}(\omega)\psi_\sigma(r)&=a \sum_{n=1}^{N} \left(\int d\mathbf{r}_1 \frac{\varphi^*_{n,\sigma}(\mathbf{r}_1)\psi_\sigma(\mathbf{r}_1)\erfc(\omega|\mathbf{r}-\mathbf{r}_1|)}{|\mathbf{r}-\mathbf{r}_1|}\right) \varphi_{n,\sigma}(\mathbf{r}) \notag \\ &\equiv a \sum_{n=1}^{N} V_{n,\psi,\sigma}^{SR}(\mathbf{r})\varphi_{n\sigma}(\mathbf{r})
\label{eq:screened_fock_sr}
\end{align}

Since $\erfc(\omega|\mathbf{r}-\mathbf{r}_1|)$ is a finite continuous function, we calculate the intergrals in Eq. \ref{eq:screened_fock_sr} with a FFT solver and a numerically optimized kernel\cite{gabay2017kernel} for the $1/r$ term, multiplied by the $\erfc$ factor. In principle, this approach can be extended to any range separated hybrid functional -- for instance, while the screening function for the BNL\cite{baer2005density} functional is different, it is also non-singular at the origin and this allowed us to use the same approach for its implementation. Thus, this scheme enables us to study large systems using any hybrid, screened hybrid and range separated functional with a moderate computational effort.

\section{Results and Discussion}
\begin{figure}[H]
\centering
\includegraphics[width=1.0\linewidth]{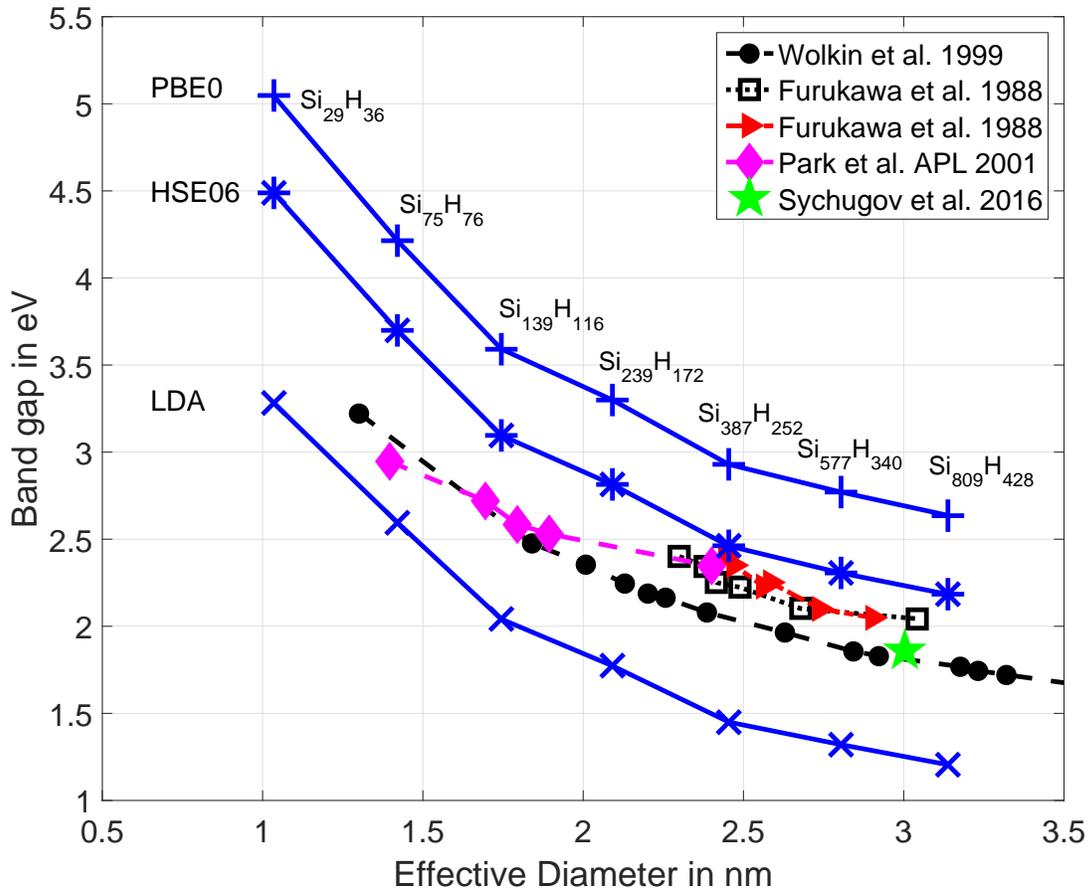}
\caption{Band gap comparison, PARSEC results are shown with blue "x" and solid line for LDA, blue asterisks and solid line for HSE06 and blue crosses and solid line for PBE0. The experimental results of Wolkin et al.\cite{wolkin1999electronic} are shown with black filled circles and dashed line, the experimental results of Park et al.\cite{park2001band} are shown in magenta filled diamonds and dashed line, the results of Furukawa et al.\cite{furukawa1988quantum} are shown with black empty squares and dotted lines and with red filled triangles and dash dot line. The result of Sychugov et al.\cite{sychugov2016single} is shown with a green filled pentagon.}
\label{fig:figure1_gap_exp}
\end{figure}

We have calculated the electronic properties of cube shaped silicon quantum dots as described by Wilson et al.\cite{wilson2014shape} who have analyzed the size and shape dependence of the band gap of quntum dots (QD) within LDA and tight binding. We have used in our calculations the QDs: \ce{Si29H36}, \ce{Si75H76}, \ce{Si139H116}, \ce{Si239H173}, \ce{Si387H252}, \ce{Si577H340}, and \ce{Si809H428}. All structures were taken from the CSIRO Nanostructure Data Bank\cite{CSIRO} and were calculated without further geometrical relaxation. We have used a real space grid spacing of $h=0.7a.u.$ and LDA based norm-conserving pseudopotentials. All caclulations were performed on a single node with 2 Intel(R) Xeon(R)  E5-2650 v2 processors with an overall number of 16 cores. Figure \ref{fig:figure1_gap_exp} shows the calculated band gap with LDA, PBE0 and HSE06 as a function of effective diameter. We estimate the effective cube edge size and effective diameter in the following way\cite{wilson2014shape} - there are $\sim$50 silicon atoms in a nm$^3$ cube, we therefore estimate $a\sim (N_{si}/50)^{(1/3)}$ where $a$ is the effective cube edge size and $N_{si}$ is the number of silicon atoms in the QD. To compare our results to spherical dots, we calculate the effective cube size for the sphere in the same way, leading to the relation $a=d\times (\pi/6)^{1/3}$, where $a$ is the effective cube edge size and $d$ is the effective sphere diameter. This way we compare dots of about the same volume, Wilson et al.\cite{wilson2014shape} have shown that the calculated gaps of spherical dots are very close to that of cubic dots with the same volume. The calculated HSE06 band gap of the largest QD that we studied, \ce{Si809H428},  is 2.18eV. This value of the gap is larger than the value of the band gap for bulk silicon calculated with HSE06 -- 1.2eV\cite{wang2014direct, jain2011reliability}. Within LDA, the calculated gap for \ce{Si809H428} is 1.23eV compared to 0.55eV of the bulk. The PBE0 band gap for the largest QD is ~2.63 eV compared to the bulk value of 1.71eV\cite{jain2011reliability}-1.85eV\cite{fuchs2007quasiparticle}.  

\begin{figure}[H]
	\centering
	\includegraphics[width=1.0\linewidth]{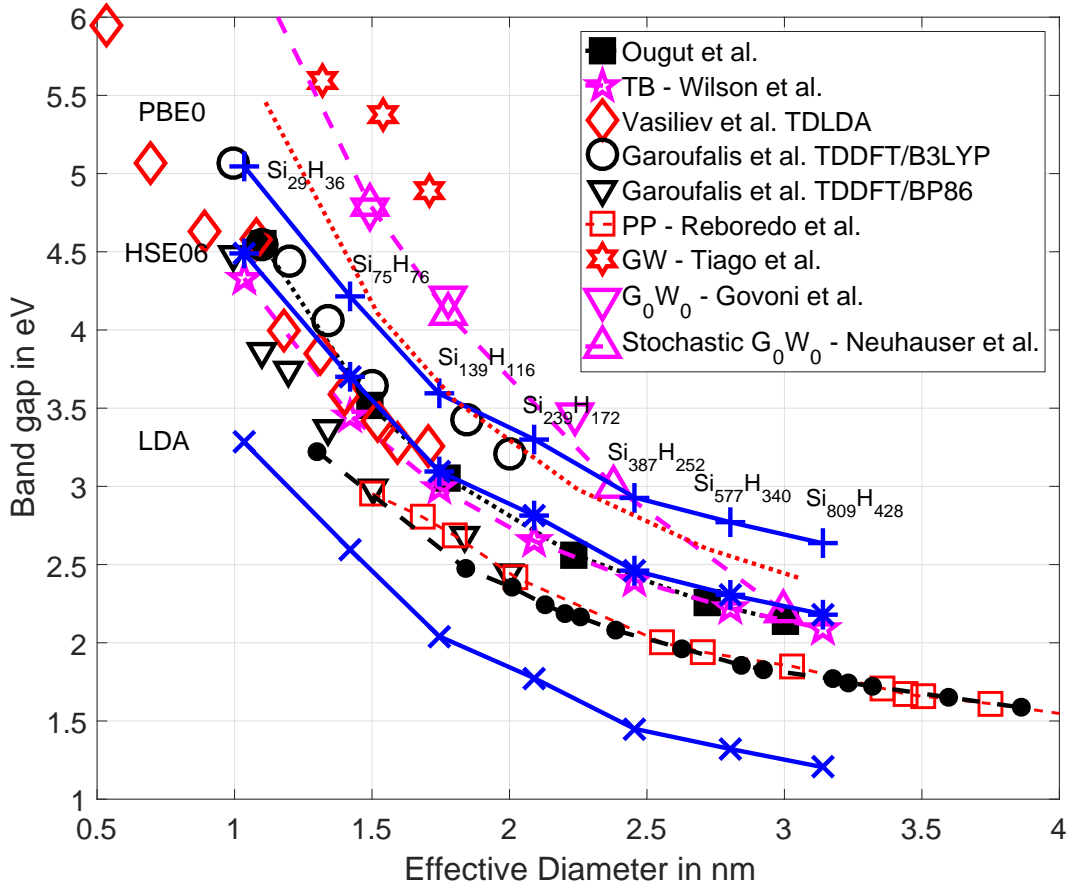}
	\caption{Band gap comparison with other theoretical calculations: Our PARSEC calculations are shown in solid blue lines, LDA with "x", HSE06 with asterisks and PBE0 with crosses. The TB results of Wilson et al.\cite{wilcoxon1999optical} are shown by magenta pentagons and dashed line, the PP results of Reboredo et al.\cite{reboredo1999excitonic} are shown by red squares and dashed line, {\"O}{\u{g}}{\"u}t et al.\cite{ougut1997quantum} optical gap is shown by black filled squares and dotted line, the TDLDA results of Vasiliev et al.\cite{vasiliev2001ab} are shown by red empty diamonds, Garoufalis et al.\cite{garoufalis2001high} TDDFT/B3LYP results are shown by black circles, Garoufalis et al.\cite{garoufalis2001high} TDDFT/BP86 results are shown with black triangles. Tiago et al.\cite{tiago2006optical} GW results are shown by red hexagons, Govoni et al.\cite{govoni2015large} $G_0W_0$ results are shown by magenta triangles pointing down, Neuhauser et al.\cite{neuhauser2014breaking} Stochastic $G_0W_0$ is shown by magenta triangles pointing up and a dashed line.  {\"O}{\u{g}}{\"u}t et al.\cite{ougut1997quantum} quasi-particle estimation is shown by a dotted red line. Wolkin et al.\cite{wolkin1999electronic} experimental results are shown with small filled black circles and a dashed line.}
	\label{fig:figure2gaptheory}
\end{figure}

Due to the large number of theoretical and experimental results we show in Figure \ref{fig:figure1_gap_exp} only our calculations and the experimental results, the comparison to other theoretical results is shown in Figure \ref{fig:figure2gaptheory}. The calculated HSE06 results for the gap are 0.1-0.6eV higher than all of the experimental optical gap measurements. The difference of the calculated HSE06 results from Wolkin et al.\cite{wolkin1999electronic} experimental results is 0.4-0.6eV with the larger values appearing for the smaller dots. The experimental results of Park et al.\cite{park2001band,park2001quantum} generally agree with those of Wolkin et al. besides a point at a diameter of 2.35nm which is higher and only 0.1eV below the HSE06 line. The results of Furukawa et al.\cite{furukawa1988quantum} are $\sim$0.2eV higher than those of Wolkin et al. and hence closer to the HSE06 line showing a difference of about 0.2eV and in some cases 0.1eV.
Recently, Sychugov et al.\cite{sychugov2016single} have performed both measurement and empirical pseudopotential theory calculations for a spherical silicon QD with a diameter of 3nm, the experimental result for the optical gap was 1.86eV while the theoretical PP result was 1.88eV. A 3nm diameter gives an interpolated value of $\sim$2.23eV on our HSE06 graph, which is 0.37eV higher.

Figure \ref{fig:figure2gaptheory} compares different theoretical calculations - Reboredo et al.{\cite{reboredo1999excitonic}} PP calculations almost agree with the experimental results of Wolkin et al.\cite{wolkin1999electronic}. To estimate the optical gap, {\"O}{\u{g}}{\"u}t et al.\cite{ougut1997quantum} have used LDA and the formula, $E_g^{opt}=E_g^{qp}-E_{coulomb}=E_{tot}(N-1)+E_{tot}(N+1)-2E_{tot}(N)-E_{coulomb}$ \ , where $E_tot$ is the LDA total energy and $E_{coulomb}$ is the estimated electron-hole interaction energy, to calculate the optical gap, their results are almost the same as our HSE06 results. Another theoretical result that agrees well with our HSE06 result is the tight-binding calculations of Wilson et al.\cite{wilson2014shape} for cubic dots. As mentioned by several authors, one should correct for the electron-hole or exciton binding energy to get the optical gap from the fundamental gap.\cite{ougut1997quantum,reboredo1999excitonic,ougut2003ab} The GW approximation can yield accurate estimation of the fundamental gap, the Bethe-Salpeter equation (BSE) and TDDFT calculations can yield good estimations for the optical gap. Vasiliev et al.\cite{vasiliev2001ab} and Garoufalis et al.\cite{garoufalis2001high} have performed TDDFT calculations of QDs with up to 147 silicon atoms - the TDDLDA results of Vasiliev et al. are close to the HSE06 line, same is true for the TDDFT/B3LYP results of Garoufalis et al., their TDDFT/BP86 are closer to experiment. Tiago and Chelickowsky have compared GW+BSE with TDLDA for silicon clusters up to \ce{Si147H100} - they got a value of 3.3eV for GW+BSE and 2.5eV with TDLDA for \ce{Si147H100} - the GW+BSE value is slightly above the HSE06 value while the the TDLDA result is almost at the experimental values of Wolkin et al.\cite{wolkin1999electronic}. Recently, Govoni et al.\cite{govoni2015large} have calculated the fundamental gap with $G_0W_0$ for QDs up to 293 silicons, Neuhauser et al.\cite{neuhauser2014breaking} have used stochastic $G_0W_0$ to calculate the fundamental gap for QDs up to 705 silicons. Interestingly, their value for the fundamental gap of \ce{Si705H300} is exactly on top of the HSE06 line and also agrees with {\"O}{\u{g}}{\"u}t et al.\cite{ougut2003ab} estimation for the optical gap. HSE06 bulk value for silicon band gap (1.2eV) is very close to the experimental gap (1.12eV), in the bulk silicon optical and fundamental gap are almost the same, but in nano size QDs the optical gap is smaller because of the exciton binding energy. HSE06 is known to yield values that are close to the optical gap in some materials, and we see here that it is above the PL experimental results for the optical gap. While there is no formal proof we can expect it to be between the fundamental and optical gap. It is therefore possible that the $G_0W_0$ results for the largest dot slightly underestimates the fundamental gap. 

There are two additional factors that are important for this comparison - the definition of diameter is not always consistent and this can cause significant shifts in some of the values of the smaller dots (less than 100 silicon). Specifically, our definition, of effective diameter, coincides only with some of the works we compared to. We show some analysis of this in the supporting information. Furthermore, we did not relax the geometry with HSE06, as LDA often leads to shorter distances, geometrical relaxation of the dots with HSE06 might yield small geometrical expansion with some reduction of the band gap, we estimate this effect as too small to explain the difference from experiment. 
 
\begin{figure}[H]
	\centering
	\begin{subfigure}{.5\textwidth}
		\centering
		\includegraphics[width=1.0\linewidth]{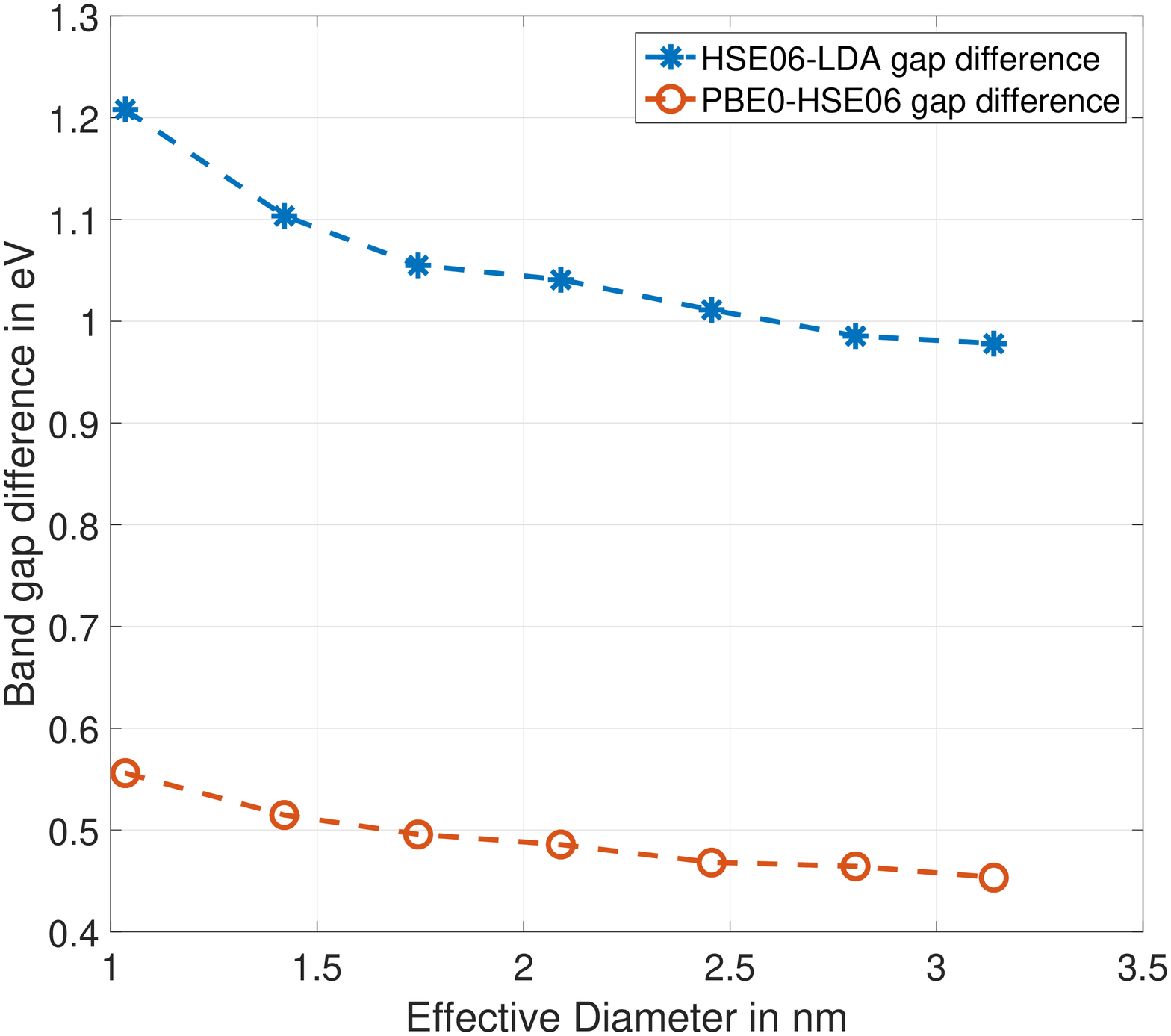}
		\caption{Band gap difference}
		\label{fig:fig2a_bg_diff}
	\end{subfigure}%
	\begin{subfigure}{.5\textwidth}
		\centering
		\includegraphics[width=1.0\linewidth]{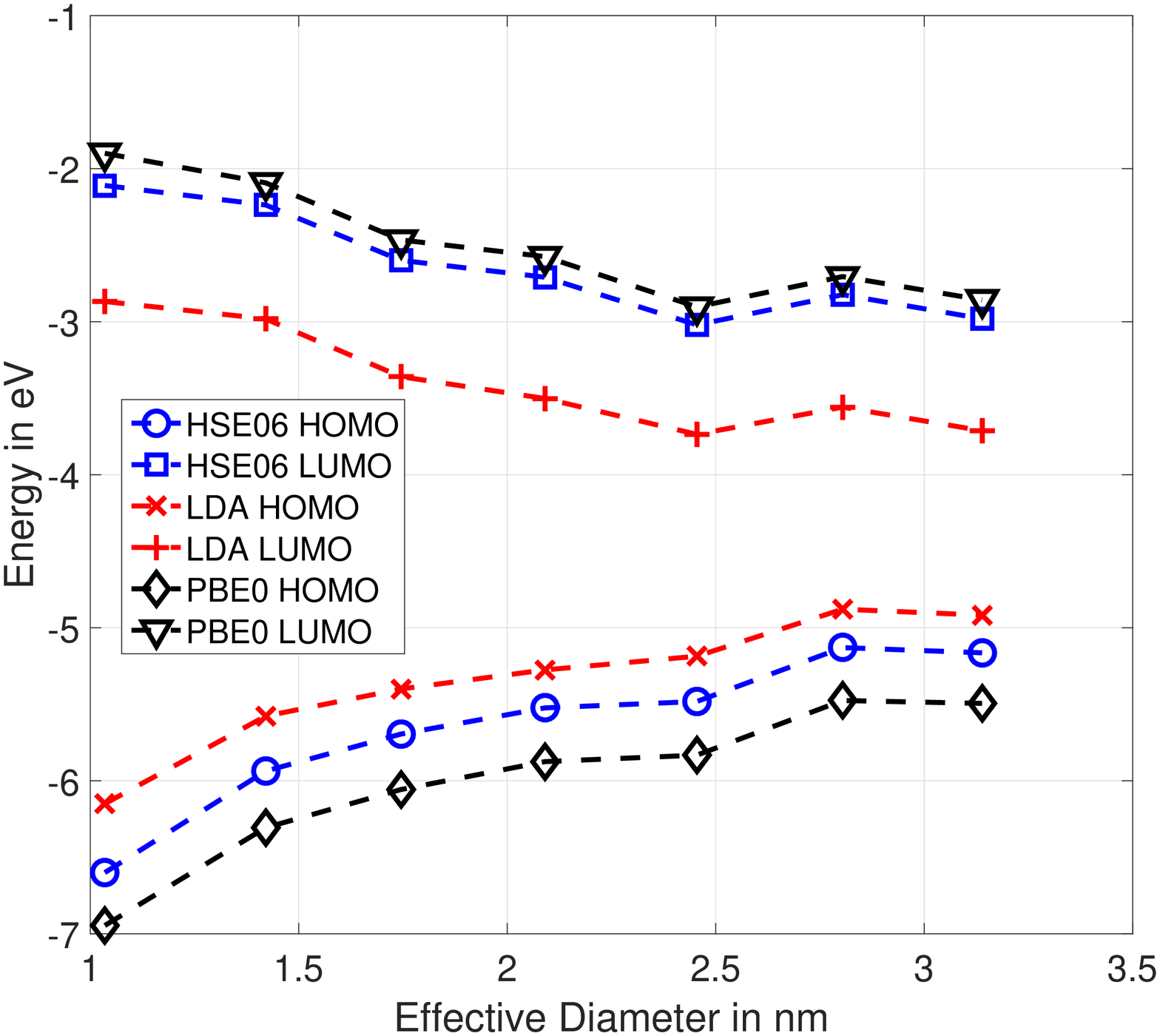}
		\caption{Band gap ratio}
		\label{fig:fig2b_bg_ratio}
	\end{subfigure}
	\caption{Comparison of HSE06 and LDA band gaps}
	\label{fig:fig2_bg_comparison}
\end{figure}

We can further compare the HSE06, PBE0 and LDA results by examination of the band gap difference between the methods (Figure \ref{fig:fig2_bg_comparison}a). 

As is evident from Figure \ref{fig:fig2_bg_comparison}a, the band gap difference between HSE06 and LDA is decreasing with size (band gap difference of $\sim 0.6$eV for the bulk). For the difference between the PBE0 and HSE06 gap, we see a smaller change with size but the same trend. In this case, the difference in the gaps is between 0.6 to 0.45 eV, compared to $\sim$0.5eV in the bulk. Comparison of the HOMO and LUMO (Lowest Unoccupied Molecule Orbital) is shown in Figure\ref{fig:fig2b_bg_ratio}. It is evident that the HSE06 LUMO is very close to the PBE0 LUMO for all sizes while the HSE06 HOMO is closer to the LDA HOMO but is almost in the middle between the PBE0 HOMO and the LDA HOMO. 

\begin{figure}[H]
\centering
\includegraphics[width=0.9\linewidth]{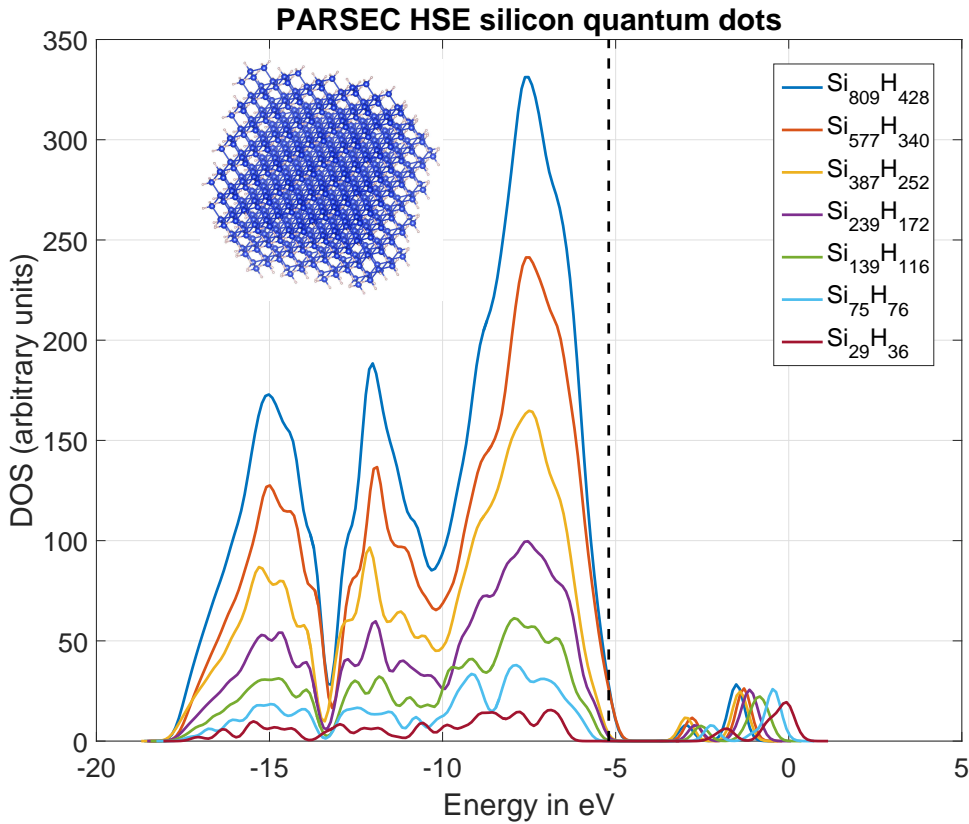}
\caption{Density of States (DOS) of occupied states for the different QD, the structure of \ce{Si809H428} is shown for illustration. The graphs are not normalized and so larger peaks are for larger dots.}
\label{fig:fig3_DOS_not_scaled}
\end{figure}

Another important indicator for the approach to bulk values is the behavior of the Density of States (DOS). This is shown in Figure \ref{fig:fig3_DOS_not_scaled}, where the convolution of eigenstates histogram with a Gaussian of 0.2eV variance has been used. As is evident from the figure, the large peak of virtual (empty) states is absent - this is because most calculations were performed with a projection of only 4 virtual states. The use of only 4 virtual states is sufficient for finding the LUMO but is clearly insufficient to represent  the full manifold of virtual states correctly. We show in the SI the effect of using more virtual states projection for the case of \ce{Si239H116}. It is interesting to note that as the size of the QDs increases,  their occupied states DOS starts to show the three peaks typical for bulk silicon. 

\begin{figure}[H]
\centering
\includegraphics[width=1.0\linewidth]{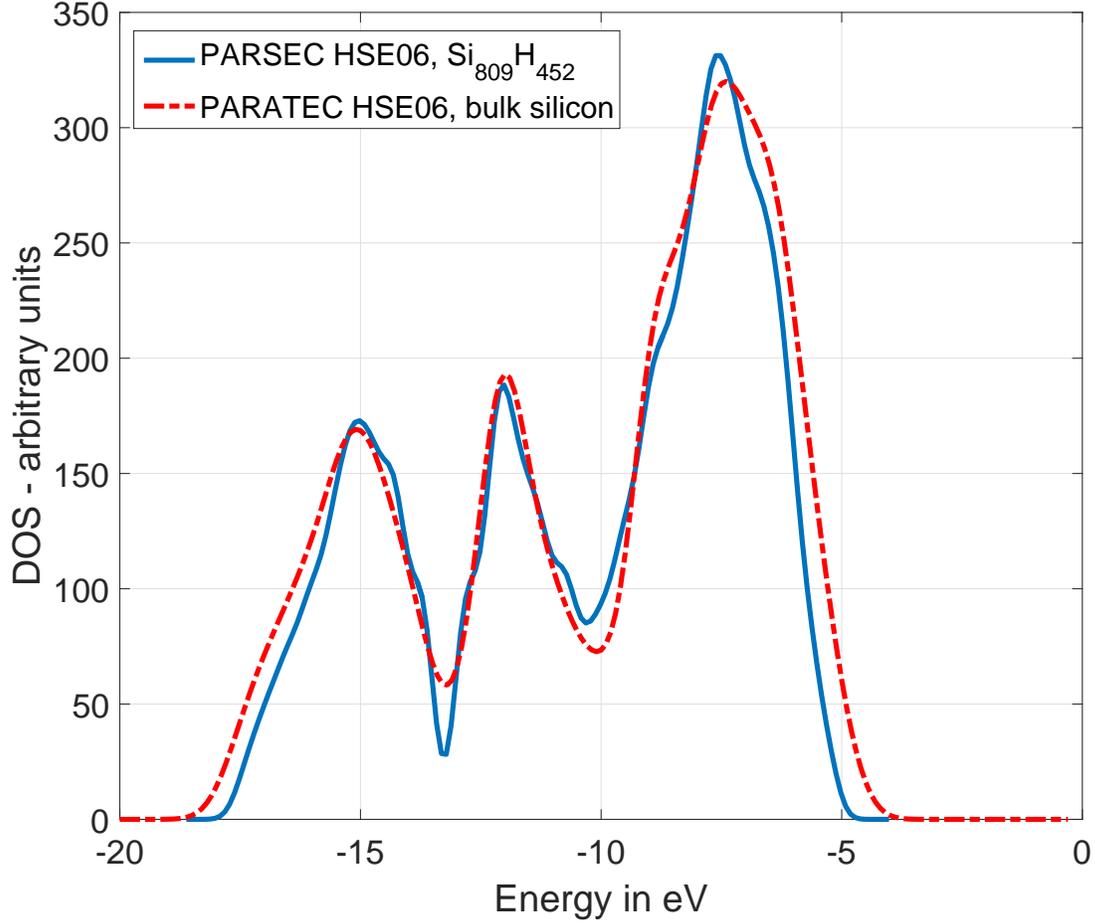}
\caption{HSE06 valence band density of states for \ce{Si809H428} (solid line) and bulk silicon (dashed line)}
\label{fig:dos_bulk_comparison_silicon}
\end{figure}

Figure \ref{fig:dos_bulk_comparison_silicon} shows the comparison of DOS of the largest QD with that of bulk Si. We calculated the DOS of valence states of bulk Si within HSE06 using the PARATEC code. A k-point mesh of $12\times 12\times 12$ and energy cutoff of 30Ry was used for the calculation. As is evident, the DOS of the QD is already very close to the bulk. The bulk graph was shifted in energy to fit best to the QD peaks. The bulk DOS
is slightly wider than the QD DOS. This can be understood by the quantum confinement effect being  still non-negligible in the largest QD studied (for instance the band gap of the QD is still not the same as the bulk).

Another property that is interesting to calculate and understand is the ionization potential (IP). The IP can be evaluated by subtracting the total energies of the neutral system (N electrons) from its cation (N-1 electrons):

\begin{equation}
IP=E_{tot}(N-1)-E_{tot}(N)
\end{equation}

We have calculated the cation total energies by removing one electron. We used spin polarized calculations for the charged system without an additional geometrical relaxation. We have calculated the IP with both LDA and HSE06, and compared our LDA results to the results of Chelikowsky et al.\cite{chelikowsky2009algorithms} who have calculated the IPs of spherical silicon QDs with LDA.

\begin{figure}[H]
\centering
\includegraphics[width=0.9\linewidth]{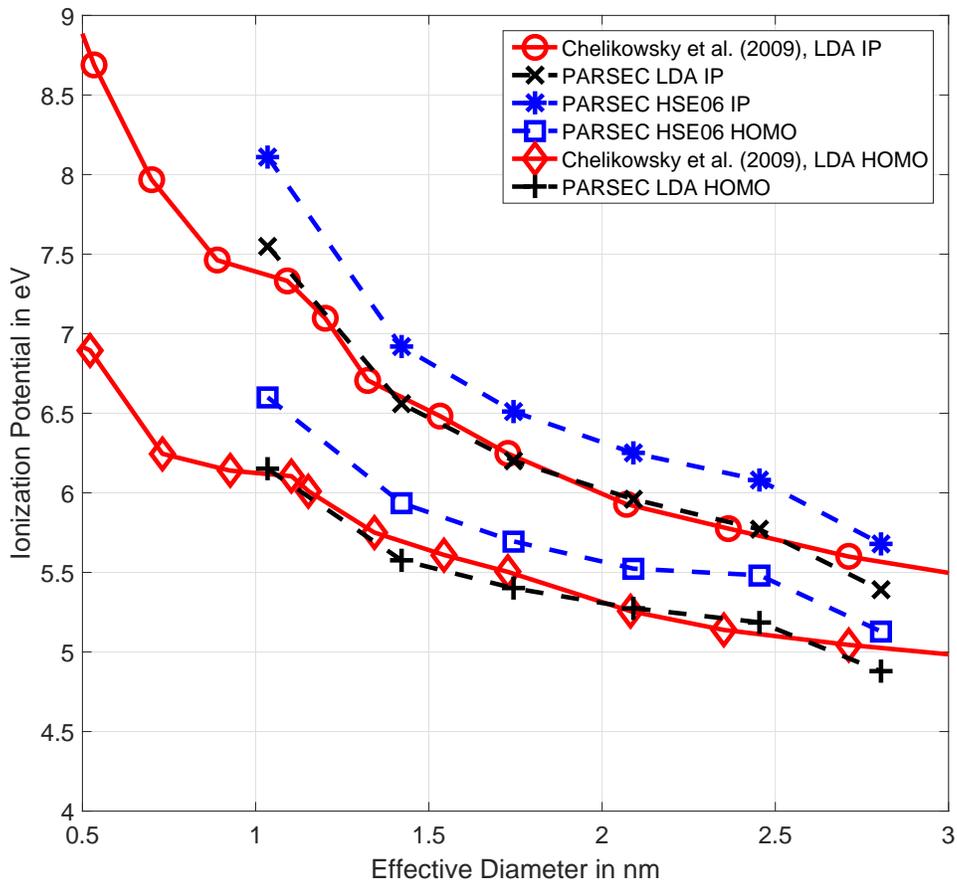}
\caption{Ionization Potential for LDA (red x symbols) and HSE06(black asterisks). We compare our LDA results to the LDA spherical dots calculations of Chelikowsky et al.\cite{chelikowsky2009algorithms} (blue circles).}
\label{fig:figure4_IP_comp}
\end{figure}

Figure \ref{fig:figure4_IP_comp} shows very good agreement, for both the HOMO and ionization potential (IP), between our LDA calculations to the calculations of Chelikowsky et al.\cite{chelikowsky2009algorithms}. The HSE06 results for both the IP and the HOMO are slightly higher than the LDA results, but the difference is relatively small compared to the band gap difference. 

\begin{figure}[H]
	\centering
	\begin{subfigure}{.5\textwidth}
		\centering
		\includegraphics[width=1.0\linewidth]{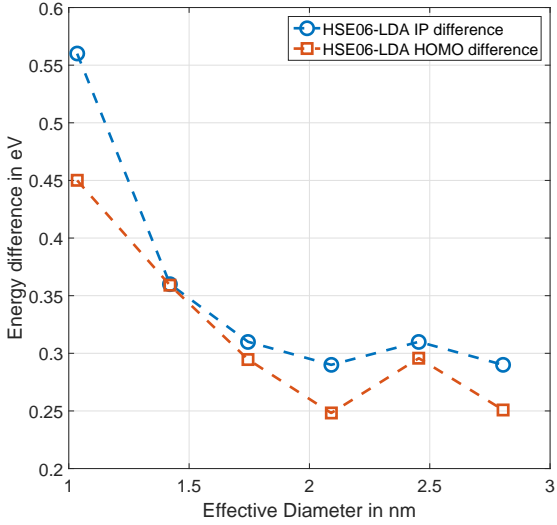}
		\caption{HSE-LDA difference}
	\end{subfigure}%
	\begin{subfigure}{.5\textwidth}
		\centering
		\includegraphics[width=1.0\linewidth]{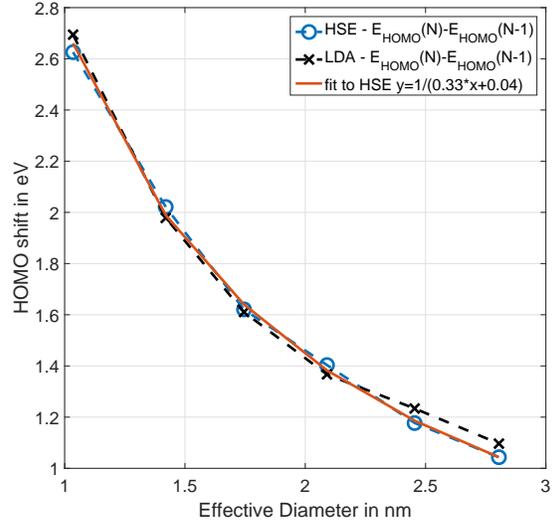}
		\caption{HOMO level shift}
	\end{subfigure}
	\caption{Comparison of HSE06 and LDA, (a) HOMO (red squares) and IP (blue circles) HSE06-LDA difference, (b) HOMO level shift with HSE06 (blue circles) and LDA (black x).}
	\label{fig:ionization_effects}
\end{figure}

Figure \ref{fig:ionization_effects}a shows that the HOMO and IP differences between HSE06 and LDA are quite similar. Both show decay with size and the difference is less than 0.3eV for systems larger than \ce{Si577H340}. Figure \ref{fig:ionization_effects}b shows the HOMO level shift after ionization  in the QDs as a function of size. It is very clear that HOMO level shift is almost identical for LDA and HSE06, we have repeated some of those calculations with PBE0 and got an almost similar shift. Furthermore, it shows a very clear decay with the dot size. This behavior can be explained by a simple electrostatic model - the HOMO is a delocalized orbital of the size of the dot, the potential of the additional charge in the ion can be modeled as a constant times $1/r$ and so we can expect that the integral of the potential times the orbital would yield a shift that behaves as $1/R$ where $R$ is proportional to the size of the dot. The fact that different levels of theory give the same value might suggest that the effect of exchange of the missing electron is negligible compared to the electrostatic effect. Following Koopman's\cite{szabo1989modern} or Janak's\cite{janak1978proof} theorems, we can expect that the HOMO level shift is the difference between the first and second ionizations. We have calculated the second ionization for dots from size 29 till 239 silicon atoms and indeed got almost the same shift - those results are shown in the SI. We could also calculate $E_{HOMO}(N-1)-E_{HOMO}(N-2)$, this shift also behaves as $E_{HOMO}(N)-E_{HOMO}(N-1)$ for the larger dots. This suggests that the difference between the third IP to the second IP is similar to the difference between the second IP to the first. For large dots this makes sense, as we can assume that the top of valence is close to degenerate and so as we ionize the main property that changes is the total charge - from neutral to +1 and then to +2, hence creating an electrostatic shift of the potential.

\section{Summary}
We have calculated the electronic properties of nanometer size silicon QD with hybrid (PBE0) and screened hybrid functionals (HSE06) up to systems of 1200 atoms. We demonstrated the trend of band gap decreasing with the size of the QD and also the difference between pure DFT calculations such as LDA and screened hybrids such as HSE06. We showed that the difference between the band gaps of the two methods is also size dependent and decreases with the size of the QD. The values we got from HSE06 are in general above the reported optical band gap experimental measurements ( differences of 0.1 to 0.6 eV ). Some TDDFT calculations give values around and below the HSE06 results. GW calculations give a fundamental gap higher than the HSE06 value for most dot sizes. It is obvious that by changing the screening parameter or the fraction of Fock exchange we can get band gaps that agree well with experiment. However, this would be an expensive semi-empirical approach and in that sense the use of PP or TB models is more reasonable. As optimally tuned range separated functionals were shown to give accurate fundamental gaps with DFT\cite{kronik2012excitation} and optical gaps with TDDFT\cite{kronik2015tddft} it would be highly interesting to evaluate them with our scheme for the larger dots.

The IP with HSE06 is higher than the LDA IP, however, the difference in IP is significantly smaller than that of the band gap and also decreases with size. We showed that the HOMO level shift and the difference between first and second ionization potentials are independent from the level of theory (HSE06, LDA and PBE0) that is used - this can be explained by electrostatic arguments and the assumption that the HOMO in the larger dots is already close to degenerate. 

The projection scheme and faster Poisson solvers make the calculation of the larger dots feasible even on a single node with 16 cores. As the parallelization of the Fock operator is easy, further acceleration can be achieved by using many compute nodes. We have also demonstrated the use of GPU to further accelerate the calculations. This opens up the possibility of combining the two approaches and use multiple GPUs to allow the study of nanostructures containing thousands of atoms using hybrid, screened hybrid and range separated functionals.

\begin{acknowledgement}
This research was supported by a grant from the United States--Israel Binational Science Foundation (BSF), Jerusalem, Israel, under BSF grant 2014426. Amir Natan thank also ISF grant 1722/13 for financial support. 
\end{acknowledgement}

\section{Appendix - Timing and GPU acceleration}
The computational bottlenecks appear at two stages: First, in the outer SCF loop, The Fock operator should be explicitly calculated for all occupied and few virtual states. This requires the calculation of Poisson integrals for all pairs of states ($\mathcal{O}(N_e^2)$) . We initially used conjugate gradient (CG) to solve the equivalent Poisson equation, with this Fock preparation stage took more than 90\% of the computational time\cite{boffi2016efficient}. To improve the speed we have switched to an FFT based Poisson integral solver\cite{gabay2017kernel} which is $\sim$10 times faster. This reduced the part of the Fock calculation to around 50\% of the time. The next time consuming stage is the projection itself. While the Projection has linear scaling with the number of electrons, $N_e$, the eigensolver diagonalization process will require more Hamiltonian calculations as the system grows and so scales also as $N_e^2$. This can be visualized in figure \ref{fig:timinglog1}: 

\begin{figure}[H]
	\centering
	\includegraphics[width=0.8\linewidth]{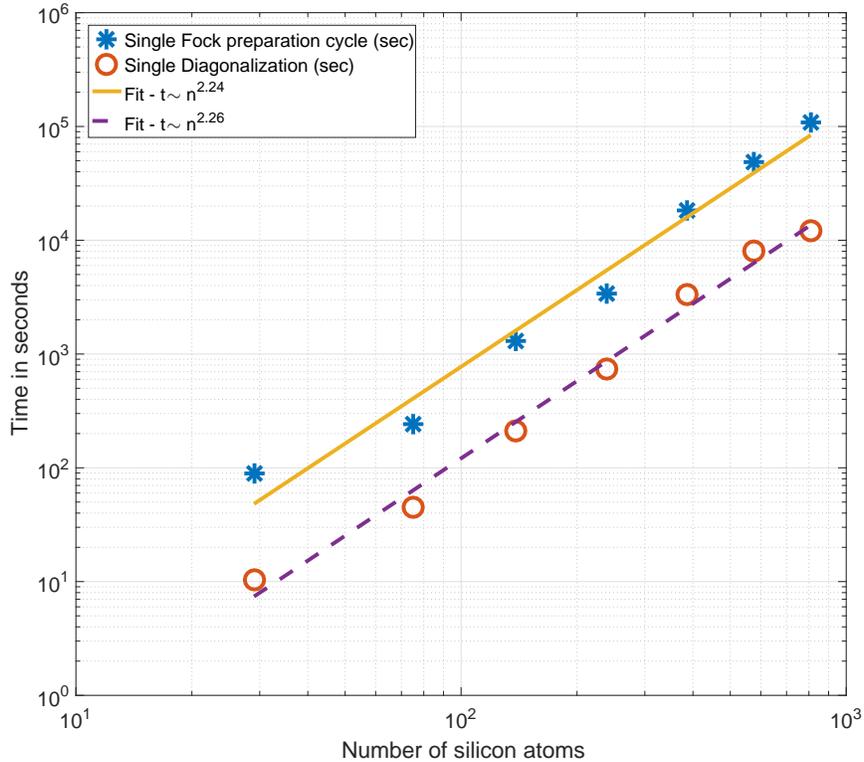}
	\caption{Timing of Fock preparation (blue asterisks and a solid fit line) and of a single diagonalization (red circles and a dashed fit line) as a function of system size.}
	\label{fig:timinglog1}
\end{figure}

We are therefore interested to accelerate with GPU the following stages:

\begin{itemize}
	\item FFT based Poisson integrals calculations
	\item Projection operation
\end{itemize} 

If the GPU memory is large enough to hold both $|\varphi_{n,\sigma}\rangle$ and $\hat{K}|\varphi_{n,\sigma}\rangle$ the projection operation can be done entirely in the GPU with a minimal data transfer cost. We have used this approach with a TESLA K40C GPU card with 2880 cuda cores, 12GB DDR4 on-board memory and $\sim$800MHz clock rate, and managed to get $\sim$4 times acceleration in the diagonalization time relative to 16 cores CPU for the \ce{Si577H340} cluster.

The GPU implementation of both the Fock preparation stage and the projection is easily integrated with the CPU code as described in algorithm \ref{alg:alg1}.

\begin{algorithm}
	\caption{Fock SCF cycle pseudocode with GPU}\label{alg:alg1}
	\begin{algorithmic}[1]
		\State perform $N_{LDA}$ cycles, $\hat{H}_{LDA}|\varphi_{n,\sigma}\rangle=\epsilon_{n,\sigma}|\varphi_{n,\sigma}\rangle$
		\State Calculate $\hat{K}[\{\varphi\}]|\varphi_{n,\sigma}\rangle$ , {\color{blue}FFT performed on GPU}
		\While{$(dfock > tol)$}
		\State $|\psi_{n,\sigma}\rangle \gets |\varphi_{n,\sigma}\rangle$
		\State $\hat{K}[\{\psi\}]|\psi_{n,\sigma}\rangle \gets \hat{K}[\{\varphi\}]|\varphi_{n,\sigma}\rangle$
		\State {\color{blue} Load $|\psi_{n,\sigma}\rangle$ and $\hat{K}[\{\psi\}]|\psi_{n,\sigma}\rangle$ to GPU}		
		\While{$(SRE>SRETOL)$}
		\State $\rho_\sigma = \sum |\varphi_{n,\sigma}|^2$
		\State Linear mixing of $V_H(\rho,r)$ with $V_H(\rho_{old},r)$
		\State Solve $\left( -\frac{\nabla^2}{2}+\hat{V}_{ps}+V_H-{\color{blue}\tilde{\hat{K}}[\{\psi\}]}\right)\varphi_{n,\sigma}=\epsilon_{n,\sigma}\varphi_{n,\sigma}$
		\State {\color{blue}$\tilde{\hat{K}}[\{\psi\}]$ is calculated with projection on GPU}
		\State Calculate Sum Residual Error (SRE) with respect to previous cycle
		\EndWhile 
		\State Calculate $\hat{K}[\{\varphi\}]|\varphi_{n,\sigma}\rangle$, {\color{blue}FFT performed on GPU}
		\State Calculate $dfock=\sum |\langle \varphi_{n,\sigma}|\hat{K}[\{\varphi\}]|\varphi_{n,\sigma}\rangle + \langle \psi_{n,\sigma}|\hat{K}[\{\psi\}]|\psi_{n,\sigma}\rangle-2\langle \varphi_{n,\sigma}|\tilde{\hat{K}}[\{\psi\}]|\varphi_{n,\sigma}\rangle|$ 
		\EndWhile
		\State Final result, energy and forces
	\end{algorithmic}
\end{algorithm}

It is possible to estimate the performance of the GPU projection by the following - suppose we $N_g$ grid points and $N_e$ orbitals. The CPU time can be given by $T_{CPU}=a_1\cdot N_g\cdot N_e$, the GPU time has a transfer time $T_{data}=a_2\cdot N_g$ which is proportional to $N_g$ and includes the data transfer and also the kinetic term that is calculated on the CPU, we can therefore write:

\begin{equation}
T_{GPU}=a_2\cdot N_g + a_3\cdot N_g\cdot N_e \Rightarrow T_{GPU}/T_{CPU}=\frac{1}{a_1}(a_2/N_e+a_3)
\end{equation}

Since the orbitals are loaded in the outer loop, during the projection we pay only the data transfer of the input guess orbital. The result is that the data transfer time becomes negligible when the number of eivenvalue is large. We have done the calculation with GPU for clusters up to \ce{Si577H340}.

\begin{figure}[H]
	\centering
	\begin{subfigure}{.5\textwidth}
		\centering
		\includegraphics[width=1.0\linewidth]{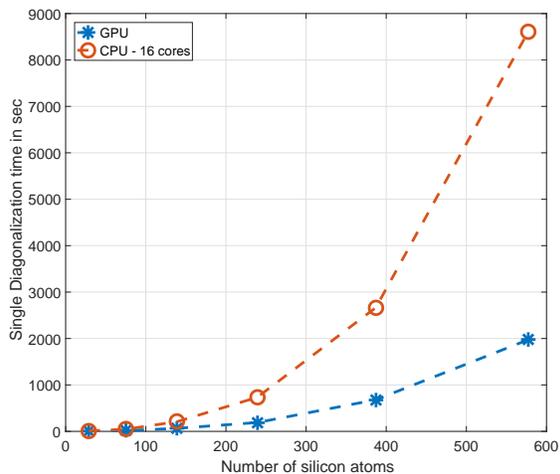}
		\caption{projection time}
		\label{fig:fig_app_a}
	\end{subfigure}%
	\begin{subfigure}{.5\textwidth}
		\centering
		\includegraphics[width=1.0\linewidth]{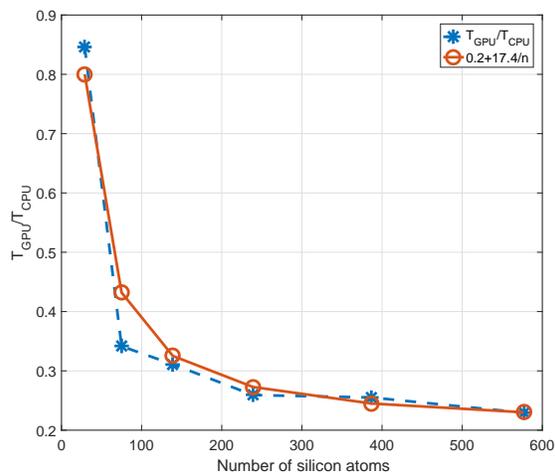}
		\caption{projection time ratio}
		\label{fig:fig_app_b}
	\end{subfigure}
	\caption{GPU vs. CPU (16 cores) diagonalization with projection timing}
	\label{fig:fig_app}
\end{figure}

\begin{suppinfo}
Additional information is found in the Supporting Information - (1) Calculation of \ce{Si239H116} Density of States with additional virtual states. (2) second ionization data (3) possible effects of diameter definition.

\end{suppinfo}

 \newcommand{\noop}[1]{}
 \providecommand{\latin}[1]{#1}
 \makeatletter
 \providecommand{\doi}
 {\begingroup\let\do\@makeother\dospecials
 	\catcode`\{=1 \catcode`\}=2 \doi@aux}
 \providecommand{\doi@aux}[1]{\endgroup\texttt{#1}}
 \makeatother
 \providecommand*\mcitethebibliography{\thebibliography}
 \csname @ifundefined\endcsname{endmcitethebibliography}
 {\let\endmcitethebibliography\endthebibliography}{}

\end{document}